\documentstyle[prd,aps,preprint]{revtex}

\input{psfig}

\begin{document}
\setcounter{page}{0}
\draft
\tighten
\title{IR Divergence and Anomalous Temperature Dependence of\\
the Condensate in the Quenched Schwinger Model}
\author{Stephan D\"urr\thanks{durr@phys.washington.edu}${}^{a,b}$ and
Stephen R. Sharpe\thanks{sharpe@phys.washington.edu}${}^{a,c}$}
\address{${}^a$University of Washington, Physics Department,
Box 351560, Seattle, WA 98195, U.S.A.\\
${}^b$Paul Scherrer Institut, Particle Theory Group, CH-5232 Villigen,
Switzerland\\
${}^c$Center for Computational Physics, University of Tsukuba, Tsukuba,
Ibaraki 305-8577, Japan\\
$\mbox{$\qquad$}$}
\vspace{3cm}
\maketitle
\begin{abstract}
The Schwinger model is used to study the artifacts of quenching in a controlled
way. The model is solved on a finite-temperature cylinder of circumference
$\beta\!=\!1/T$ with bag-inspired local boundary conditions at the two ends
$x^1\!=\!0$ and $x^1\!=\!L$ which break the $\gamma_5$-invariance and thus play
the role of a small quark mass. The quenched chiral condensate is found to
diverge exponentially as $L\to\infty$, and to diverge (rather than melt as for
$N_{\rm f}\!\geq\!1$) if the high-temperature limit $\beta\!\to\!0$ is
taken at finite box-length $L$. We comment on the generalization of our results
to the massive quenched theory, arguing that the condensate is finite as
$L\to\infty$ and proportional to $1/m$ up to logarithms.
\end{abstract}
\vfill

Preprint numbers: UW-PT/98-19, PSI-PR-00-04, UTCCP-P-55, hep-lat/9902007


\clearpage

%%%%%%%%%%%%%%%%%%%%%%%%%%%%%%%%%%%%%%%%%%%%%%%%%%%%%%%%%%%%%%%%%%%%%%%%%%%%%%%

% A useful Journal macro
\def\J#1#2#3#4{{#1} {\bf #2} (#4), #3}
% Some useful journal names
\def\NPB{{\em Nucl.Phys.} B}
\def\NPP{\em Nucl. Phys. (Proc. Suppl.)}
\def\PLB{{\em Phys. Lett.} B}
\def\PRD{{\em Phys. Rev.} D}
\def\PRL{\em Phys. Rev. Lett.}
\def\CMP{\em Commun. Math. Phys.}
\def\AOP{\em Ann. Phys.}
\def\JPA{{\em J. Phys.} A}
\def\JPQ{\em J. Physique (France)}
\def\HPA{\em Helv. Phys. Acta}

\newcommand{\pad}{\partial}
\newcommand{\pas}{\partial\!\!\!/}
\newcommand{\Dsl}{D\!\!\!\!/\,}
\newcommand{\Psl}{P\!\!\!\!/\;\!}
\newcommand{\Nf}{N_{\rm f}}
\newcommand{\hqu}{\hbar}
\newcommand{\ovr}{\over} 
\newcommand{\hal}{{1\ovr2}}
\newcommand{\til}{\tilde}
\newcommand{\pri}{^\prime}
\renewcommand{\dag}{^\dagger}
\newcommand{\<}{\langle}
\renewcommand{\>}{\rangle}
\newcommand{\lan}{\langle}
\newcommand{\gaf}{\gamma_5}
\newcommand{\lap}{\triangle}
\newcommand{\trc}{\rm tr}
\newcommand{\al}{\alpha}
\newcommand{\be}{\beta}
\newcommand{\ga}{\gamma}
\newcommand{\de}{\delta}
\newcommand{\ep}{\epsilon}
\newcommand{\ve}{\varepsilon}
\newcommand{\ze}{\zeta}
\newcommand{\et}{\eta}
\newcommand{\th}{\theta}
\newcommand{\vt}{\vartheta}
\newcommand{\io}{\iota}
\newcommand{\ka}{\kappa}
\newcommand{\la}{\lambda}
\newcommand{\rh}{\rho}
\renewcommand{\vr}{\varrho}
\newcommand{\si}{\sigma}
\newcommand{\ta}{\tau}
\newcommand{\up}{\upsilon}
\newcommand{\ph}{\phi}
\newcommand{\vp}{\varphi}
\newcommand{\ch}{\chi}
\newcommand{\ps}{\psi}
\newcommand{\om}{\omega}
\newcommand{\psb}{\overline{\psi}}
\newcommand{\etb}{\overline{\eta}}
\newcommand{\psd}{\psi^{\dagger}}
\newcommand{\chd}{\chi^{\dagger}}
\newcommand{\etd}{\eta^{\dagger}}
\newcommand{\etp}{\eta^{\prime}}
\newcommand{\rch}{{\rm cosh}}
\newcommand{\rsh}{{\rm sinh}}
\newcommand{\beq}{\begin{equation}}
\newcommand{\eeq}{\end{equation}}
\newcommand{\bdm}{\begin{displaymath}}
\newcommand{\edm}{\end{displaymath}}
\newcommand{\bea}{\begin{eqnarray}}
\newcommand{\eea}{\end{eqnarray}}

%%%%%%%%%%%%%%%%%%%%%%%%%%%%%%%%%%%%%%%%%%%%%%%%%%%%%%%%%%%%%%%%%%%%%%%%%%%%%%%

\section{Introduction}

%%%%%%%%%%%%%%%%%%%%%%%%%%%%%%%%%%%%%%%%%%%%%%%%%%%%%%%%%%%%%%%%%%%%%%%%%%%%%%%

\noindent
Simulations of lattice QCD often use the quenched approximation in which the
quark determinant is replaced by a constant \cite{QuenchedSimulations}.
While it is clear that the quenched theory is unphysical \cite{Morel}, it is
less certain what pathologies are introduced in Euclidean correlation
functions.
Considerations based on quenched chiral perturbation theory suggest that the
chiral limit is singular \cite{QuenchedTheory}.
For example, the quark condensate is predicted to diverge in this limit:
$\<\psb\psi\>\propto m^{-\delta}$, with $\delta\approx 0.1$
\footnote{This divergence is not related to that due to exact zero modes in 
topologically non-trivial backgrounds, an effect which is suppressed in the
infinite volume limit \cite{LeutwylerSmilga}.}.
There is some numerical evidence supporting the predictions of quenched chiral
perturbation theory, but it is not yet definitive \cite{NumericalEvidence}.

Given this situation, it is interesting to explore the effect of quenching in
a completely controlled environment where calculations can be done 
analytically.
In this paper we undertake such an investigation for the Schwinger model (QED
in two dimensions with $\Nf$ massless fermions \cite{SMorg}).
This has been frequently used as a toy model for QCD(4), since it shares the
properties of confinement and dynamical mass generation \cite{SMcon}.
In particular, we have calculated the chiral condensate in the massless
quenched theory with chiral symmetry breaking introduced by the spatial
boundary conditions.
The chirality violating parameter is $1/L$ ($L$ being the box length) instead
of the quark mass $m$.
We also introduce thermal boundary conditions in the Euclidean time direction.
In this way we can formally work with the massless theory throughout (the
finite spatial and temporal extent of the manifold act as an infrared
regulator), and maintain full analytic control.
In particular, the result for the theory with positive $\Nf$ quantized with
these boundary conditions can be analytically continued to the quenched limit,
$\Nf=0$
\footnote{In the presence of an infrared regulator, this is equivalent to
directly calculating correlation functions in the quenched approximation, i.e.\
without the determinant.}.
We are able to calculate the condensate in the quenched theory at any spatial
location as a function of $L$ and the inverse temperature $\beta$.

A more straightforward approach would be to quantize the massive theory with
an arbitrary number of flavors in a finite box, and then take the limits
$\Nf\to0$, $L\to\infty$ and $m\to0$ sequentially (in this particular order).
The difficulty here is that even in the case of the massive QED(2) only an
approximate solution exists \cite{SMmassive}.
Since analytical continuation in $\Nf$ is, in general, arbitrarily sensitive
to small variations in input data, one is on more secure ground using an exact
solution.
The drawback in our approach of breaking the chiral symmetry by boundary
conditions is that the last two limits mentioned above ($L\to\infty$ and
$m\to0$) are effectively being taken simultaneously.

There have been several previous investigations of the quenched Schwinger
model, both analytical and numerical.
Casher and Neuberger argue that the spectral density of the Dirac operator at
zero eigenvalue is non-zero~\cite{CasherNeuberger} --- an observation which, 
if we were in four dimensions, 
would indicate spontaneous chiral symmetry breaking.
Introducing an ad-hoc infrared regulator mass $\mu_{\rm IR}$, Carson and Kenway
\cite{CarsonKenway} and Grandou~\cite{Grandou} conclude, using bosonization
methods, that the quenched condensate actually diverges exponentially as the
regulator is removed:
\beq
\<\psd\psi\> \sim
\mu_{\rm IR} \exp\left({e^2 \over 2 \pi\mu_{\rm IR}^2}\right)
\;.
\label{eq:Grandou}
\eeq
Smilga reaches a similar conclusion based on an analysis of the eigenvalue
spectrum of the Dirac operator~\cite{Smilga}. 
Our calculation shows that the result~(\ref{eq:Grandou}) is just a specific
form of the generic combination ``power-law fall-off times exponential
divergence'', different versions of which we find as exact asymptotic results
when the boundaries are sent to infinity in various ways.
Our approach uses a finite box with specific chirality breaking boundary
conditions rather than an ad-hoc IR-regulator and allows for a generalization
to finite temperatures.

The following section summarizes previous work on the Schwinger model with
chirality breaking boundary conditions \cite{WiDu,DuWi,Du}.
We then describe the analytic continuation to $\Nf=0$, and study its behavior
in the interesting limits.
In sec.~\ref{sec:discussion} we elucidate the origin of the IR divergences
that we find.
Section~\ref{sec:massive} is devoted to a more speculative discussion
of the IR divergences in the massive quenched theory.
We conclude with a summary of our results.

%%%%%%%%%%%%%%%%%%%%%%%%%%%%%%%%%%%%%%%%%%%%%%%%%%%%%%%%%%%%%%%%%%%%%%%%%%%%%%%

\section{Chiral Condensate with Arbitrary Number of Flavors}

%%%%%%%%%%%%%%%%%%%%%%%%%%%%%%%%%%%%%%%%%%%%%%%%%%%%%%%%%%%%%%%%%%%%%%%%%%%%%%%

\subsection{Manifold Parameters and Abbreviations}

\noindent
The Euclidean Schwinger model (massless QED in $d=2$ dimensions)
\bea
S[A,\psd,\ps]&=&S_B[A]+S_F[A,\psd,\ps]
\nonumber
\\
S_B={1\ovr4}\int\limits_M  F_{\mu\nu} F_{\mu\nu}\quad&,&\quad
S_F=\sum\limits_{n=1}^{N_{\!f}} \int\limits_{M}\psd_n \Dsl\;\ps_n
\label{mfsm.1}
\eea
is studied on the manifold
\beq
M=[0,\be]\times[0,L]\quad\ni\quad(x^0,x^1)
\label{pro.1}
\eeq
with volume $V=\be L$.
In Euclidean time direction, the fields $A$ and $\psi$ are periodic and
antiperiodic respectively with period $\beta$. Hence $x^0=0$ and $x^0=\beta$
are identified (up to a sign) and the manifold is a cylinder.
At the two spatial ends of the cylinder (i.e.\ at $x^1=0$ and $x^1=L$) some
specific chirality-breaking (XB-) boundary-conditions (which will be discussed
below) are imposed. We use the one-flavor Schwinger mass
\beq
\mu\;:\;\equiv\;{|e|\ovr\sqrt{\pi}}
\label{schwingermass}
\eeq
to simplify our notation, and introduce the dimensionless inverse temperature
and box-length
\beq
\si\equiv\mu\be\qquad\qquad\la\equiv\mu L
\label{natuni}
\eeq
as well as the dimensionless volume, aspect ratio, and spatial position
\beq
\up=\si\la\qquad,\qquad\tau={\si\ovr2\la}\qquad,\qquad\xi={x^1\ovr L}
\;.
\label{abbrev}
\eeq

%%%%%%%%%%%%%%%%%%%%%%%%%%%%%%%%%%%%%%%%%%%%%%%%%%%%%%%%%%%%%%%%%%%%%%%%%%%%%%%

\subsection{Chirality Breaking Boundary Conditions}

\noindent
The proposal to study both QCD and the Schwinger model with chirality-breaking
boundary-conditions goes back to Ref.~\cite{HrBa}.
The XB-boundary conditions can be motivated by requiring the operator $i\Dsl$
to be symmetric under the scalar product $(\ch,\ps):=\int\,\chd\ps\,d^2x$,
which leads to the condition that the surface integral $\oint\chd
n\!\!\!\slash\,\ps\,ds$ vanishes, where $n\!\!\!\slash=\!n_\mu\ga_\mu$ with
$n_\mu$ denoting the outward oriented normal on the boundary.
Imposing local linear boundary conditions which ensure this requirement amounts
to having $\chd n\!\!\!\slash\,\ps=0\;$ on the boundary for each pair.
A sufficient condition to guarantee this is to require all modes to obey
$\ps=B\ps$ on the boundary, where the boundary operator $B$ (which is
understood to act as the identity in flavor space) has to satisfy
$B\dag n\!\!\!\slash B=-n\!\!\!\slash$ and $B^2=1$.
In \cite{WiDu,DuWi,Du} the one-parameter family of solutions
\beq
B\equiv B_\th:\equiv i\gaf e^{\th\gaf}n\!\!\!\slash
\label{boundaryoperator}
\eeq
was chosen, supplemented by suitable boundary conditions for the gauge-field.
The $\gaf$ invariance of the theory is broken for all $\th$, thus making the
$\Nf$-flavor theory invariant under $SU(\Nf)_V$ instead of $SU(\Nf)_L\,\times
SU(\Nf)_R$. 
In physical terms, these boundary conditions prevent the $U(1)$-current
from leaking through the boundary, since $j\cdot n=\psd n\!\!\!\slash\,\ps=0$
on $\pad M$.
There are considerable differences regarding the spectrum of the Dirac
operator in the theory with XB-boundary-conditions \cite{WiDu,DuWi} as
compared to the theory on the torus \cite{SaWi}:
\begin{itemize}
\item[-]
The Dirac operator has a discrete real spectrum which is {\em asymmetric}
with respect to zero.
\item[-]
The spectrum is empty at zero, i.e.\ the Dirac operator has {\em no zero
modes}.
\item[-]
The instanton number $q=e/(4\pi)\cdot\int\ep_{\mu\nu}F_{\mu\nu}=
e/(2\pi)\cdot\int E\;\in {\it{\bf R}}\/$ is {\em not quantized}.
\end{itemize}
The first property already indicates that we are away from the usual
Atiyah-Patodi-Singer index-theorem situation.
The fact that the second property is fulfilled implies that the generating
functional for the fermions in a given gauge-field background
\beq
Z_F[A,\etd,\et]\;=\;
{1\ovr N}\int \prod_{i=1}^{\Nf} D\psd_{(i)} D\ps^{}_{(i)}\ 
e^{-\sum\int\psd_{(i)}\Dsl\ps_{(i)}+
\sum\int\psd_{(i)}\et_{(i)}+\sum\int\etd_{(i)}\ps_{(i)}}
\label{mfsm.2}
\eeq
does indeed simplify to the textbook formula
\beq
Z_F[A,\etd,\et]\;=\;
\Big({{\det}_\th(\Dsl)\ovr{\det}_\th(\,\pas\,)}\Big)^{\Nf}\
e^{\,\int\etd(\Dsl)^{-1}\et}
\label{mfsm.3}
\eeq
from which the one-flavor condensate in a given background [no sum over $(i)$]
\beq
\<\psd_{(i)}(x)P_\pm\ps_{(i)}(x)\>\!\!{{}\atop\mbox{\small $A$}}={1\ovr Z_F}\;
{\de^2\ovr\de\et^{}_{(i)\,\pm}(x)\;\de\etd_{(i)\,\pm}(x)}\,Z_F\;
\Big\vert_{\et^{}_\pm=\etd_\pm=0}
\label{mfsm.4}
\eeq
is computed.
Here and in the following $P_\pm\!=\!1/2\cdot(1\pm\gaf)$ denotes the projector
on the two chiralities, where $\gaf\!=\!\mbox{diag}(1,-1)$ in the chiral
representation of the Dirac Clifford algebra.

%%%%%%%%%%%%%%%%%%%%%%%%%%%%%%%%%%%%%%%%%%%%%%%%%%%%%%%%%%%%%%%%%%%%%%%%%%%%%%%

\subsection{General Result at Arbitrary Points}

\noindent
The condensate in the $\Nf$-flavor Schwinger Model on a finite-temperature
cylinder with the XB-boundary-conditions (\ref{boundaryoperator}) at the two
spatial ends was found to read \cite{DuWi,Du}
\bea
{\<\psd P_{\pm}\ps\>\ovr\mu}&=&
\pm{e^{\pm\th\cdot\rch(\la\sqrt{\Nf}(1-2\xi)/2)/\rch(\la\sqrt{\Nf}/2)}
\ovr4\la}\cdot
\nonumber
\\
&{}&
\sum_{n\in Z}(-1)^n
{\sin(\pi\xi)\rch(\pi n\ta)\ovr\sin^2(\pi\xi)+\rsh^2(\pi n\ta)}\cdot
{\int\limits_{-1/2}^{1/2}dc\;\cos(2\pi nc)\,\theta_3^{\Nf}(c,i\tau)\ovr
\int\limits_{-1/2}^{1/2}dc\;\,\theta_3^{\Nf}(c,i\tau)}\cdot
\nonumber
\\
&{}&
\exp\Big\{{1\ovr\Nf}\sum\limits_{n\geq1}
\Big(1-\cos(2\pi n\xi)\Big)
\Big({{\rm coth}(\pi n\tau)\ovr n}-
(n\rightarrow\sqrt{n^2\!+\Nf(\la/\pi)^2\,})\Big)\Big\}
\label{2.1}
\\
\nonumber
\\
{\<\psd P_{\pm}\ps\>\ovr\mu}&=&
\pm{e^{\pm\th\cdot\rch(\la\sqrt{\Nf}(1-2\xi)/2)/\rch(\la\sqrt{\Nf}/2)}
\ovr2\si}\cdot
\nonumber
\\
&{}&
\sum_{m\in Z}(-1)^m
{1\ovr\rsh(\pi(m\!+\!\xi)/\ta)}\cdot
{\int\limits_{-1/2}^{1/2}\!dc\;\rch(2\pi(m\!+\!\xi)c/\ta)\,
e^{-\Nf\pi c^2/\ta}\;\theta_3^{\Nf}(ic/\tau,i/\tau)\ovr
\int\limits_{-1/2}^{1/2}\!dc\;\,e^{-\Nf\pi c^2/\tau}\;
\theta_3^{\Nf}(ic/\tau,i/\tau)}\cdot\!\!
\nonumber
\\
&{}&
\exp\Big\{{\pi\ovr\Nf}
\Big({\xi(1\!-\!\xi)\ovr\tau}+
{\rch(\la\sqrt{\Nf}(1\!-\!2\xi))-\rch(\la\sqrt{\Nf})
\ovr\si\sqrt{\Nf}\;\rsh(\la\sqrt{\Nf})}\Big)\Big\}\cdot
\label{2.2}
\\
&{}&
\exp\Big\{{1\ovr\Nf}\sum\limits_{m\geq1}
{\rch(\pi m/\tau)-\rch(\pi m(1\!-\!2\xi)/\tau)
\ovr m\;\rsh(\pi m/\tau)}-
(m\rightarrow\sqrt{m^2\!+\Nf(\si/2\pi)^2\,})\Big\}
\;,
\nonumber
\eea
where the integration variable $c$ represents the harmonic piece in the Hodge
decomposition of the gauge potential, and the elliptic function $\th_3$ is
defined as
\beq
\theta_3(u,\om)=\sum_{n\in Z}e^{2\pi inu}q^{n^2}=
1+2\sum_{n\geq1}\cos(2n\pi u)q^{n^2}\qquad\quad(q\equiv e^{i\pi\om})
\label{thet.4}
\eeq
for the parameters $\om=i\ta$ and $\om=i/\ta$ (giving real nome $q\in]0,1[$).
The two forms (\ref{2.1}, \ref{2.2}) are identical for any finite $\si$ and
$\la$, but they enjoy excellent convergence properties in the regimes
$\ta\gg 1$ and $\ta\ll 1$, respectively ($\ta=\si/2\la$).

%%%%%%%%%%%%%%%%%%%%%%%%%%%%%%%%%%%%%%%%%%%%%%%%%%%%%%%%%%%%%%%%%%%%%%%%%%%%%%%

\section{Quenched Chiral Condensate}

%%%%%%%%%%%%%%%%%%%%%%%%%%%%%%%%%%%%%%%%%%%%%%%%%%%%%%%%%%%%%%%%%%%%%%%%%%%%%%%

\subsection{Quenched Chiral Condensate at Arbitrary Points}

\noindent
This paper is based on the observation that -- even though certain expressions
within formulas (\ref{2.1}, \ref{2.2}) diverge as $\Nf\to 0$ (for fixed $\si,
\la$) -- the product of all factors stays finite:
\bea
{\<\psd P_{\pm}\ps\>\ovr\mu}&=&
\pm{e^{\pm\th}\ovr4\la}
{1\ovr\sin(\pi\xi)}\cdot
\exp\Big\{{\la^2\ovr2\pi}\sum\limits_{n\geq1}
\Big(1\!-\!\cos(2\pi n\xi)\Big)
\Big({{\rm coth}(\pi n\tau)\ovr\pi n^3}+
{\ta\ovr n^2\rsh^2(\pi n\ta) }\Big)
\Big\}
\label{2.3}
\\
\nonumber
\\
{\<\psd P_{\pm}\ps\>\ovr\mu}&=&
\pm{e^{\pm\th}\ovr2\si}
{\ta\ovr\sin(\pi\xi)}\cdot
\exp\Big\{
{\pi\si^2\ovr12\ta^3}\xi^2(1-\xi)^2
\Big\}\cdot
\nonumber
\\
&{}&
\exp\Big\{{\si^2\ovr8\pi\ta}
\sum_{m\geq1}{1\ovr m^2\rsh^2(\pi m/\ta)}
\Big(1-\xi\rch({2\pi m(1\!-\!\xi)\ovr\ta})-
(1\!-\!\xi)\rch({2\pi m\xi\ovr\ta})\Big)\cdot
\nonumber
\\
&{}&\!
\exp\Big\{{\si^2\ovr8\pi^2}
\sum_{m\geq1}{1\ovr m^3\rsh(\pi m/\ta)}
\Big(\rch({\pi m\ovr\ta})-\rch({\pi m(1\!-\!2\xi)\ovr\ta})\Big)
\;,
\label{2.4}
\eea
where the low- and high-temperature forms (\ref{2.3}, \ref{2.4}) are
identical for any finite $\si,\la$.

%%%%%%%%%%%%%%%%%%%%%%%%%%%%%%%%%%%%%%%%%%%%%%%%%%%%%%%%%%%%%%%%%%%%%%%%%%%%%%%

\subsection{Specialization to the Quenched Case at Midpoints}

\noindent
To study chiral symmetry breaking with XB-boundary conditions,
one has to determine whether the condensate 
in the bulk of the cylinder survives when the boundaries are sent to
infinity. We choose to consider the midpoint ($\xi\!=\!1/2$), 
and also to set the boundary parameter $\th$ to zero, for which the
resulting expressions are most simple. Other values of $\xi$ and $\th$ 
lead to the same conclusions.
Specializing eqs. (\ref{2.3}, \ref{2.4}) to $\xi\!=\!1/2$ and $\th=0$,
and rewriting them in terms of
the variables $\up=\si\la$ and $\ta=\si/2\la$, we find
\bea
{\<\psd P_{\pm}\ps\>\ovr\mu}&=&
\pm{\sqrt{\ta\,}\ovr2^{3/2}\sqrt{\up\,}}\cdot
\exp\Big\{{\up\ovr2\pi\ta}\sum\limits_{n\geq0}
\Big({{\rm coth}(\pi(2n\!+\!1)\ta)\ovr\pi(2n\!+\!1)^3}+
{\ta\ovr(2n\!+\!1)^2\rsh^2(\pi(2n\!+\!1)\ta)}\Big)
\Big\}
\label{2.5}
\\
\nonumber
\\
{\<\psd P_{\pm}\ps\>\ovr\mu}&=&
\pm{\sqrt{\ta\,}\ovr2^{3/2}\sqrt{\up\,}}\cdot
\exp\Big\{
{\pi\up\ovr96\ta^2}
+{\up\ta\ovr4\pi}
\sum\limits_{m\geq1}
\Big({{\rm tanh}(\pi m/2\ta)\ovr\pi m^3}
-{1\ovr2 m^2\ta\rch^2(\pi m/2\ta)}\Big)
\Big\}
\;,
\label{2.6}
\eea
These two expressions enjoy
excellent convergence properties in the regimes $\ta\gg 1$ and $\ta\ll 1$,
respectively.

%%%%%%%%%%%%%%%%%%%%%%%%%%%%%%%%%%%%%%%%%%%%%%%%%%%%%%%%%%%%%%%%%%%%%%%%%%%%%%%

\subsection{Asymptotic Expansions for $\ta\gg1$ and $\ta\ll1$}

\noindent
From (\ref{2.5}, \ref{2.6}) the condensate is seen to behave asymptotically
like
\bea
{\<\psd P_{\pm}\ps\>\ovr\mu}
&\sim&
\pm{1\ovr4\la}
\exp\Big\{{\la^2\ovr\pi^2}
\sum\limits_{n\geq0}
{1\ovr(2n\!+\!1)^3}\Big\}
=
\pm{1\ovr4\la}\exp\Big\{{7\ze(3)\la^2\ovr8\pi^2}\Big\}
\quad\!\qquad\qquad\;({\mbox{\small low temp.}\atop\mbox{\small fixed $L$}})
\label{2.7}
\\
\nonumber
\\
{\<\psd P_{\pm}\ps\>\ovr\mu}
&\sim&
\pm{\ta\ovr2\si}
\exp\Big\{{\si^2\ovr4\pi^2\ta^2}
\sum\limits_{n\geq0}
{1\ovr(2n\!+\!1)^3}\Big\}
=
\pm{\ta\ovr2\si}\exp\Big\{{7\ze(3)\si^2\ovr32\pi^2\ta^2}\Big\}
\qquad\qquad\;\;\,({\mbox{\small $L\to0$}\atop\mbox{\small fixed $T$}})
\label{2.8}
\\
\nonumber
\\
{\<\psd P_{\pm}\ps\>\ovr\mu}
&\sim&
\pm{\ta\ovr2\si}
\exp\Big\{
{\pi\si^2\ovr192\ta^3}
+{\si^2\ovr8\pi^2}
\sum\limits_{m\geq1}
{1\ovr m^3}\Big\}
=
\pm{e^{\ze(3)\si^2/8\pi^2}\ta\ovr2\si}
\exp\Big\{{\pi\si^2\ovr192\ta^3}\Big\}
\quad\;({\mbox{\small $L\to\infty$}\atop\mbox{\small fixed $T$}})
\label{2.9}
\\
\nonumber
\\
{\<\psd P_{\pm}\ps\>\ovr\mu}
&\sim&
\pm{1\ovr4\la}
\exp\Big\{
{\pi\la^2\ovr48\ta}
+{\la^2\ta^2\ovr2\pi^2}
\sum\limits_{m\geq1}
{1\ovr m^3}\Big\}
\sim
\pm{1\ovr4\la}
\exp\Big\{{\pi\la^2\ovr48\ta}\Big\}
\quad\!\!\qquad\qquad\,({\mbox{\small high temp.}\atop\mbox{\small fixed $L$}})
\label{2.10}
\eea
where only the leading term is kept.

%%%%%%%%%%%%%%%%%%%%%%%%%%%%%%%%%%%%%%%%%%%%%%%%%%%%%%%%%%%%%%%%%%%%%%%%%%%%%%%

\subsection{Divergence of the quenched chiral condensate in the chiral limit}

\noindent
As noted in the introduction, we are particularly interested in the behavior of
the condensate in the chiral limit.
The condensate is properly 
defined by first taking the infinite volume limit and
then sending the chirality breaking source to zero.
In our framework these two limits are taken simultaneously by sending
$\la\to\infty$.
This means that in general we do not know {\em a priori} whether the source is
``strong enough'' to lead to symmetry breaking when $\la\to\infty$.
For example, in QCD(4) if $m\to0$ and $L\to\infty$ such that
$m V \to 0$ in the limit, with $V$ the Euclidean volume, then chiral
symmetry breaking will not occur.
But if we find a non-zero limiting value for the condensate, we conclude
{\em a posteriori} that the symmetry breaking source is strong enough.
Indeed, we shall see that in the (massless) quenched Schwinger model the
condensate diverges as $\la\to\infty$ for any temperature.

\begin{figure}[t]
\vspace{-2.5cm}
\psfig{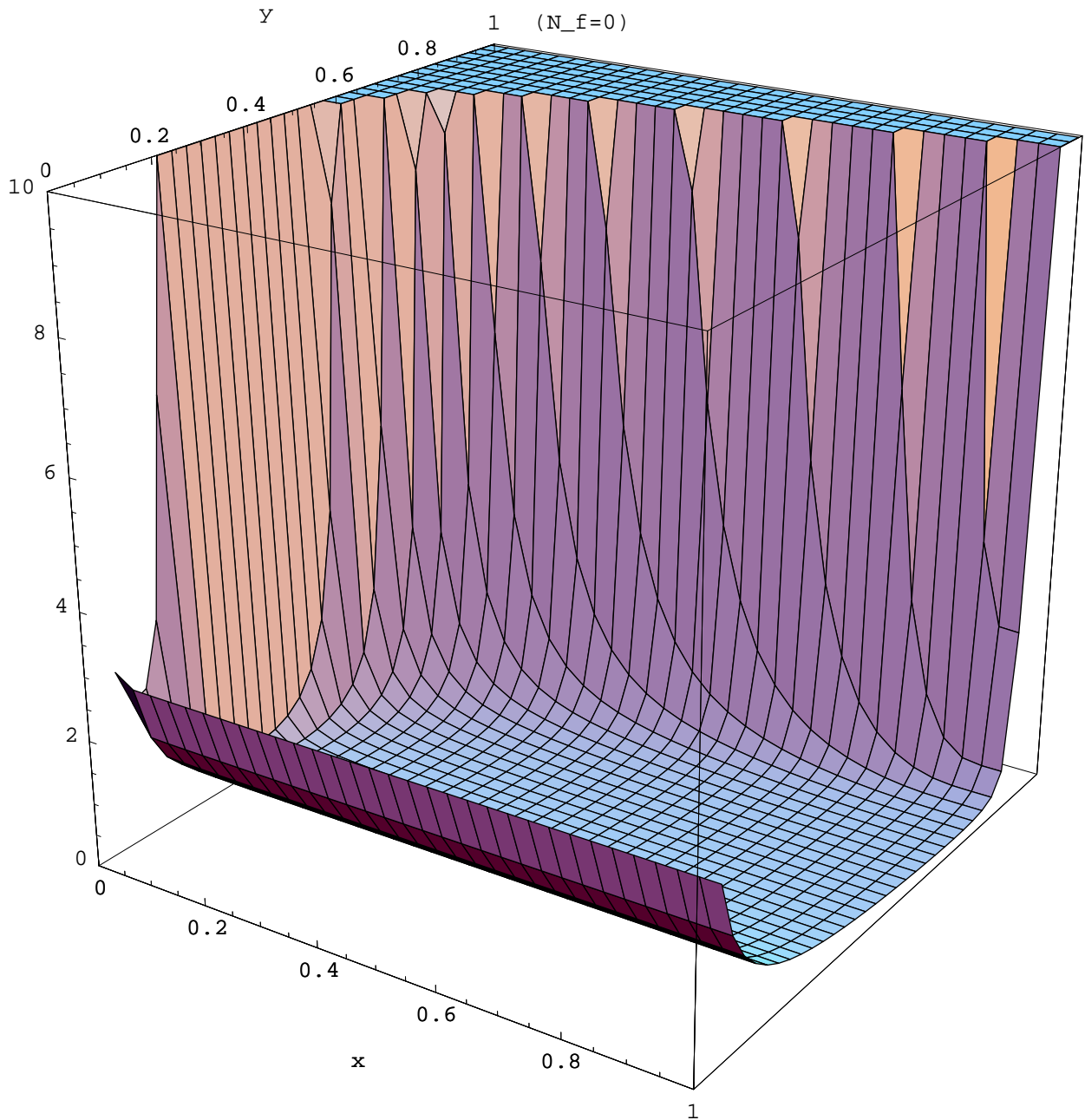}
\vspace{-1.5cm}
\caption{The quenched condensate at the midpoint 
as a function of spatial box size and inverse temperature,
using eqs. (\ref{2.14}, \ref{2.15}).
The axes have been rescaled so that the entire range of values is shown:
$x\!=\!2/\pi\cdot{\rm arctan}(\si)$ and $y\!=\!2/\pi\cdot{\rm arctan}(\la)$,
where $\si\!=\!\mu\be$ and $\la=\mu L$. Unlike the other divergencies, 
the one near $y\!=\!0$ is unphysical as it is an artifact of the boundary
conditions. The border of the surface at $x\!=\!1$ represents the (massless)
quenched Schwinger model at zero temperature with our boundary conditions.}
\label{fig1}
\end{figure}

\begin{enumerate}
\item
To study the limit of infinite box-length at zero temperature we first perform
the limit of zero temperature in (\ref{2.7}) which gives the exact result
\beq
\lim_{\si\to\infty}
{\<\psd P_{\pm}\ps\>\ovr\mu}\Big\vert_{\la\;{\rm fixed}}
=
\pm{1\ovr4\la}\cdot\exp\Big\{{7\ze(3)\ovr8\pi^2}\la^2\Big\}
\;.
\label{2.11}
\eeq
Clearly the infinite volume limit cannot be taken.
\item
To study the limit of infinite box-length at fixed (non-zero, non-infinite)
temperature we use (\ref{2.9}) from which we see that the condensate now
diverges like
\beq
{\<\psd P_{\pm}\ps\>\ovr\mu}
\sim
\pm{e^{\ze(3)\si^2/8\pi^2}\ovr4\la}
\exp\Big\{{\pi\ovr24\si}\la^3\Big\}
\qquad({\la\gg1\atop\si\mbox{ fixed}})
\label{2.12}
\eeq
which is even more virulent than in the zero-temperature case (\ref{2.11}).
\item
To study the limit of infinite box-length at infinite temperature we first
focus on the high-temperature limit at finite box-length. From (\ref{2.10})
we find the result
\beq
{\<\psd P_{\pm}\ps\>\ovr\mu}
\sim
\pm{1\ovr4\la}\exp\Big\{{\pi\la^3\ovr24}{1\ovr\si}\Big\}
\qquad({\si\ll1\atop\la\mbox{ fixed}})
\label{2.13}
\eeq
which diverges at infinite temperature even at finite box-length.
\item
Finally one may check whether the results are significantly altered if the
limit of infinite box-length is taken while simultaneously lowering or raising
the temperature in such a way that either the aspect ratio $\ta=\si/2\la$ or
the box-volume $\up=\si\la$ stays constant.
\newline
Under $\up\to\infty$ at fixed $\ta$ the condensate is seen from (\ref{2.5})
to diverge like
\beq
{\<\psd P_{\pm}\ps\>\ovr\mu}
\sim
\pm\mbox{const }{1\ovr\sqrt{\up}}\exp\Big\{\mbox{const }\up\Big\}\qquad.
\label{2.14}
\eeq
Under $\ta\to0$ at fixed $\up$ the condensate is seen from (\ref{2.6})
to diverge like
\beq
{\<\psd P_{\pm}\ps\>\ovr\mu}
\sim
\pm\mbox{const }\sqrt{\ta}\exp\Big\{\mbox{const }{1\ovr\ta^2}\Big\}\qquad.
\label{2.15}
\eeq
\end{enumerate}
Our findings are illustrated in figure \ref{fig1}.

%%%%%%%%%%%%%%%%%%%%%%%%%%%%%%%%%%%%%%%%%%%%%%%%%%%%%%%%%%%%%%%%%%%%%%%%%%%%%%%

\section{Origin of infrared divergences}
\label{sec:discussion}

%%%%%%%%%%%%%%%%%%%%%%%%%%%%%%%%%%%%%%%%%%%%%%%%%%%%%%%%%%%%%%%%%%%%%%%%%%%%%%%

\bigskip

It is interesting to elucidate the origin of the IR divergences
in the quenched theory.
Mathematically, things can be traced back to the fact that in two dimensions
the gauge-field can be decomposed into a coexact and an exact piece ($A_\mu=
-\ep_{\mu\nu}\pad_\nu\ph+\pad_\mu\ch\;$ plus, with certain boundary conditions,
a topological and a harmonic contribution) and the functional determinant (if
present) can be computed by integrating the anomaly.
As a consequence, the full effective action in the $\Nf$-flavor Schwinger model
is found to read (up to contributions from the topological and harmonic parts)
\beq
S_{\Nf}[\ph]={1\ovr2e^2}\int\ph(\lap^2-\Nf{e^2\ovr\pi}\lap)\ph\;dx\;.
\label{conc1}
\eeq
Going back to (\ref{mfsm.2},~\ref{mfsm.3},~\ref{mfsm.4}) and using
decomposition properties of the massless Dirac operator in a given gauge-field
(which are specific to two dimensions), the condensate with our boundary
conditions and for arbitrary $\Nf\!\geq\!0$ is found to factorize
\bea
\<\psd(x)P_\pm\ps(x)\>&=&
{\int dc\, D\ph\ \ S_\th(x,x)_{\pm\!\pm}\
e^{-\Gamma_{\th,N_{\!f}}[c,\ph]}
\ovr
\int dc\, D\ph\ \
e^{-\Gamma_{\th,N_{\!f}}[c,\ph]}}
\nonumber
\\
{}&=&
{\int dc\ \til S_\th(x,x)_{\pm\!\pm}\ e^{-N_{\!f}\Gamma(c)}
\ovr\int dc\ \ e^{-N_{\!f}\Gamma(c)}}
\;\cdot\;
{\int\ D\ph\ e^{\mp 2\ph(x)-\Gamma_{\th,N_{\!f}}[\ph]}
\ovr\int\ D\ph\ \ e^{-\Gamma_{\th,N_{\!f}}[\ph]}}
\;.
\label{conc2}
\eea
Here $\til S_\th(x,x)_{\pm\!\pm}$ denotes the diagonal entries of the Green
function of $i\pas$ (obeying XB-boundary-conditions)
and $\Gamma_{\th,N_{\!f}}[\ph]$ is just $S_{\Nf}[\ph]$, 
up to a term which vanishes for $\th\!=\!0$
(for further details the reader is referred to
\cite{DuWi,Du}).
The first integral in (\ref{conc2}) 
(which is over a $c$-number) is trivial in the quenched theory.
The second integral (which is an integration over all $\ph$-fields which are
periodic over $\be$ and satisfy Dirichlet boundary conditions at $x^1\!=\!0$
and $x^1\!=\!L$) can be done exactly for all $\Nf$:
\beq
{\int\ D\ph\ e^{\mp 2\ph(x)-S_{N_{\!f}}[\ph]}
\ovr\int\ D\ph\ \ e^{-S_{N_{\!f}}[\ph]}}
=
\exp\Big\{2e^2\<x\|{1\ovr\lap^2-\Nf{e^2\ovr\pi}\lap}\|x\>\Big\}\;.
\label{conc3}
\eeq
We stress that the factorization property (\ref{conc2}) reflects itself in
the form of our analytical results --- both in the dynamical case
[eqns.~(\ref{2.1},~\ref{2.2})] and the quenched case
[eqns.~(\ref{2.3},~\ref{2.4})]:
All expressions consist of a prefactor [the first two lines in
eqns.~(\ref{2.1},~\ref{2.2}) or the factor $1/(4\la\sin(\pi\xi))$ in
eqns.~(\ref{2.3},~\ref{2.4})] and a product of exponentials.
In both cases the prefactor corresponds to the first factor in (\ref{conc2}),
whereas the exponentials represent the second factor in
(\ref{conc2}), as expressed in (\ref{conc3}).

The important point here is that in the quenched case the IR divergences are
due entirely to the second factor in (\ref{conc2}), i.e.\ due to the divergence
of (\ref{conc3}). 
Indeed, as our explicit expressions show, the first factor in~(\ref{conc2})
vanishes as a power-law in the limit of infinite box-length in both quenched
and full theories.
This is as expected, since the leading effect of the chirally asymmetric
boundary conditions on the free fermion propagator, $\tilde S(x,x)$, is
proportional to the propagator from $x$ to the boundary and back, and this
propagator falls off since the Dirac operator has a mass-dimension lower than
the dimension of the manifold.

By contrast, the operator $\lap^2-\Nf\mu^2\lap$,
which appears in the expression (\ref{conc3}),
has a mass-dimension higher than the dimension of the manifold.
Its propagator 
\beq
{\cal G}(x-y;N_f) = \<x\|{1\ovr\lap^2-\Nf\mu^2\lap}\|y\>
\;,
\label{eq:calGdef}
\eeq
thus grows with distance, and one has the possibility
that the boundary conditions become more influential as the spatial length is
increased. 
To understand this in more detail, it is
useful to have explicit expressions for the requisite Green functions.
In the quenched theory one finds a quadratic increase
\beq
{\cal G}(x-y;N_f=0) = %\<x\|{1\ovr\lap^2}\|y\>=
{1\ovr8\pi}(x\!-\!y)^2\log(\mu|x\!-\!y|)-
{1\ovr8\pi}(x\!-\!y)^2\;,
\label{conc6}
\eeq
while for $\Nf\!\geq\!1$ the result increases only logarithmically
\beq
{\cal G}(x-y;N_f\ge 1) =
%\<x\|{1\ovr\lap^2-\Nf\mu^2\lap}\|y\>=
{1\ovr\Nf\mu^2}\ 
\Big(-{1\ovr2\pi}\log(\Nf^{1/2}\mu|x\!-\!y|)-
{1\ovr2\pi}K_0(\Nf^{1/2}\mu|x\!-\!y|)\;\Big)
\;.
\label{conc5}
\eeq
The less rapid increase in the unquenched case turns out,
as shown by the explicit results, to be insufficient to lead to
an IR divergence in the condensate.

%\footnote{It is, however, UV-finite. This is explicitly true for the quenched
%theory [see (\ref{conc6})], and follows in the full theory since the diagonal
%element is the difference of two Green functions with the same singularity
%structure [see (\ref{conc5})].}

One can push the argument a little further by imagining using images to
enforce the boundary conditions. 
One cannot actually repeat our calculation in this way, because the sum over
the two-fold infinite set of images does not converge, since contributions grow
as the images become more distant --- both in quenched and full theories.
But one can use images to obtain the pattern of IR divergences observed in
our results by considering the simpler geometry of a semi-infinite spatial
extent at zero temperature: at $x^1\!=\!0$ we apply XB-boundary conditions
whereas for $x^1\!\to\!\infty$ we require $L^2$-integrability.
The analogous quantity to that which we have calculated is the condensate at a
distance $L/2$ from the left boundary, and the important issue is how it
behaves as $L\to\infty$.
In this new geometry we need only a single image, and it is the Green function
between the original position and its image  which dominates for large $L$.
The asymptotic forms are thus
\beq
\exp\Big\{2e^2\<x\|{1\ovr\lap^2-\Nf{e^2\ovr\pi}\lap}\|x\>\Big\}
= 
\left\{
{
\exp\Big\{\mbox{const}\cdot \la^2 \log(\la)\Big\} \quad (N_f=0)
\atop
\exp\Big\{\mbox{const}\cdot\log(\la)\Big\} \hfill (N_f \ge 1)
}
\right.
\label{conc7}
\eeq
where we use $\la=L |e| /\sqrt\pi$ instead of $L$
to aid the comparison with our previous results.
We expect that our results should have a similar form, but with different
constants, due to the other mirror copies present in our geometry.

This expectation turns out to be correct, up to logarithmic accuracy.
For the quenched case, $\Nf\!=\!0$, our exact result at zero  temperature,
eqn.~(\ref{2.11}), takes the form $\exp(0.107 \la^2)$.
It thus matches eqn.~(\ref{conc7}), except that the logarithm is absent.
For $\Nf\!=\!1$ the form (\ref{conc7}) works with ``const'' equal to 1, which
is exactly what is needed to cancel the asymptotic $1/L$-behavior of the first
factor in (\ref{conc2}) and to reproduce the well-known value for 
the condensate
of the one-flavor Schwinger model in the limit $L\!\to\!\infty$ \cite{DuWi}.
For $\Nf\!=\!2$ the ``const'' is found to be $1/2$, which is not sufficient to
cancel the asymptotic $1/L$-behavior of the first factor in (\ref{conc2}).
Consequently the two-flavor condensate vanishes as $1/\sqrt{L}$ when the model
is quantized with XB-boundary conditions \cite{Du}. 

This completes our attempt to make sense out of the functional form of our
analytical results.
While is interesting to see that the IR-sickness of the quenched Schwinger
model can be traced back to the {\em long-distance behavior} of the Green
function $\<x|\lap^{-2}|y\>$, it is clear that the argument is specific to
two dimensions.

%%%%%%%%%%%%%%%%%%%%%%%%%%%%%%%%%%%%%%%%%%%%%%%%%%%%%%%%%%%%%%%%%%%%%%%%%%%%%%%

\section{Speculations on the massive theory}
\label{sec:massive}

%%%%%%%%%%%%%%%%%%%%%%%%%%%%%%%%%%%%%%%%%%%%%%%%%%%%%%%%%%%%%%%%%%%%%%%%%%%%%%%

Having an analytical result for the (massless) quenched Schwinger model it is
interesting 
to consider the IR behavior of the corresponding massive model, 
quenched QED(2).
In particular, we would like to know what happens if one disentangles
the limits $L\to\infty, m\to 0$ and takes them in the standard order: first
$L\to\infty$ and then $m\to0$.
In QCD(4) these two limits do not commute: the standard order leads to
a non-zero condensate, whereas $\<\psb\ps\>$ vanishes when $m\to0$ is taken
at any finite $L$.
In quenched QED(2) knowledge concerning these two situations is asymmetric.
Taking the massless limit first, analytic results are obtained and, as we have
seen, one cannot take the $L\to\infty$ limit when $m\!=\!0$.
The result of taking the limits in the standard order is, 
however, controversial.
There is no exact analytical expression for the condensate for $m\ne0$,
and so one must resort to approximate methods.
Reference~\cite{Grandou} argues that sending $L\to\infty$ at fixed $m$
yields a finite result for the condensate, and that this result remains
finite as $m\to 0$. This is what is found explicitly in the approximate
calculation of Ref.~\cite{GuerinFried}.
On the other hand 
Ref.~\cite{Steele} proposes a power-law divergence as $L\to\infty$.
In the following we analyze the situation using two approaches.
First, we expand in powers of the quark mass,
and find that each coefficient diverges exponentially as the volume
is sent to infinity.
Second, we generalize the approach of Smilga~\cite{Smilga}, 
and argue that there is
in fact no IR divergence at finite $m$, and that the IR divergences in our
first approach are artifacts of expanding about $m=0$.
The IR divergence does, however, reappear as a power law divergence
when $m\to0$, in disagreement with Refs.~\cite{Grandou,GuerinFried}.

The derivatives of the condensate with respect to $m$ at $m\!=\!0$ are higher
correlation functions in the massless theory, and can be computed analytically.
Note that, since our boundary conditions break chiral symmetry, there is no
reason to expect non-analyticity about $m\!=\!0$ for finite volume.
%We expect, for our XB-boundary conditions, that quantities 
%are analytic about $m\!=\!0$ at finite $L$, since chiral symmetry is
%broken at the boundaries and the point $m\!=\!0$ is not special.
%By contrast, for chirally symmetric boundary conditions (e.g. on a torus)
%there is no reason to expect any quantity to be analytic around $m\!=\!0$.

We start from the form of
the condensate in the massive quenched theory 
\beq
\<\psd P_+\ps\>(x)=
{\int dc\;D\ph\;\;
(S_{\th,m}(x,x)_{++}\!-\!\til S_{\th,m}(x,x)_{++}\!+\!\til S_{\th,0}(x,x)_{++})
\;\;
e^{-{1\ovr2e^2}\int\ph\lap^2\ph}
\ovr
\int dc\;D\ph\;\;
e^{-{1\ovr2e^2}\int\ph\lap^2\ph}}
\;,
\label{conc9}
\eeq
where $S_{\th,m}$ is the inverse of $\Dsl\!+\!m$ with XB boundary conditions,
and $\til S_{\th,m}$ its counterpart at zero gauge-coupling.
The subtractions are chosen such that (\ref{conc9}) vanishes in the limit
$e\to0$ at fixed $m$, while still reducing to (\ref{conc2}) when $m\to0$ at
fixed $e$. In addition, they make the expression UV-finite.
To get a sense of the general term in the expansion about $m\!=\!0$,
we consider the first derivative,
\bea
{d\<\psd P_+\ps\>(x)\ovr dm}\Big\vert_{m\!=\!0}
&=&
\hal\sum_{\pm}\int\!dy\,
\int\!dc\;\til S_\th(x,y)_{+ \pm}\til S_\th(y,x)_{\pm +}\;\times
\nonumber
\\
&{}&\qquad\qquad\;\;
\Bigg({\int\ D\ph\ e^{-2\ph(x)\mp 2\ph(y)-{1\ovr2e^2}\int\ph\lap^2\ph}
\ovr\int\ D\ph\ \ e^{-{1\ovr2e^2}\int\ph\lap^2\ph}}-1\Bigg)
\nonumber
\\
&=&
\hal\sum_{\pm}\int\!dy\,
\int\!dc\;\til S_\th(x,y)_{+ \pm}\til S_\th(y,x)_{\pm +}\;\times
\label{conc10}
\\
&{}&\qquad\qquad\;\;
\Bigg(\!\exp\Big\{2e^2\Big(
\mbox{$
\<x|{1\ovr\lap^2}|x\>\!\pm\!\<x|{1\ovr\lap^2}|y\>\!\pm\!
\<y|{1\ovr\lap^2}|x\>\!+\!\<y|{1\ovr\lap^2}|y\>
$}
\Big)\Big\}-1\Bigg)
\;.
\nonumber
\eea
Here $y$ runs over the entire manifold, the $c$-integration is the average over
a $c$-number valued harmonic part of the gauge field (necessary since the
cylinder is not simply connected~\cite{DuWi}), and the sum consists of two
terms (i.e.\ the $\pm$ are either all $+$ or all $-$).
The explicit forms of $\til S_\th$ (the free
massless propagator subject to the XB-boundary-conditions) and
$\<x|\lap^{-2}|y\>$ are given in \cite{DuWi}.
What is important here is that, for geometrically fixed $x$ and $y$,
$\<x|\lap^{-2}|y\>$ diverges as $L^2$.
Thus, for at least one of the choices of sign, the second factor in
(\ref{conc10}) diverges as $e^{L^2}$.
This dominates the power-law fall-off of the first factor.
Thus there is at least a set of $y$-values with nonzero measure for which the
combined argument of the $y$-integration in (\ref{conc10}) is IR-divergent
\footnote{On this point the unquenched theory differs significantly: Each Green
function diverges only logarithmically with $L$, and the overall divergence of
the second factor is compensated ($\Nf\!=\!1$) or overwhelmed ($\Nf\!\geq\!2$)
by the fall-off of the $c$-integral.}.
Similar arguments can be made for the higher derivatives of the condensate.

We conclude that each coefficient in the expansion of the condensate 
about $m\!=\!0$ is highly IR divergent.
Of course, this does not necessarily imply that the 
quenched condensate diverges when $V\to\infty$ at fixed $m$,
because the limits may not commute.
For example if the quenched condensate had the following schematic
dependence on $m$ and $V$
\beq
\<\psd P_\pm \ps\>\simeq{1\ovr m+\exp(-V)}
\qquad,
\label{mixedcond0}
\eeq
then it would be finite as $V\to\infty$ at fixed $m$,
while derivatives at $m=0$ are all IR divergent.
In the following we argue that a form like eq.~(\ref{mixedcond0})
actually holds.

We use the line of reasoning initiated by Smilga~\cite{Smilga},
generalized to the quenched case.
We briefly recapitulate his argument.
The key result (valid for $\Nf\!\geq\!1$ and for $\Nf\!=\!0$) is
\beq
\<\sum_n {1\ovr\la_n^2}\>=
{1\ovr 2\pi^2}\int\,d^2x\,d^2y\,{1\ovr(x-y)^2}
\exp\{2e^2({\cal G}(0;N_f)-{\cal G}(x-y;N_f))\}
\;.
\label{gengreen}
\eeq
This relates the sum of the inverse-squares of the eigenvalues
\footnote{In the remainder of this section we use $\lambda$ to denote
eigenvalues and not the dimensionless box length.}
of the Dirac operator to the Green function appearing in ~(\ref{conc6},
\ref{conc5}).
As $V\to\infty$ the leading behavior of the r.h.s. of (\ref{gengreen}) is
determined by the asymptotic form of ${\cal G}$.
For unquenched theories one finds, using (\ref{conc5}),
\beq
\<\sum_n {1\ovr\la_n^2}\>\propto
V^{(\Nf+1)/\Nf}\qquad\qquad(\Nf\!\ge\!1)
\quad,
\label{sumrorig}
\eeq
with missing dimensions provided by powers of $e$.
This implies that the characteristic size of the lowest eigenvalues is
\beq
\la_c\propto
V^{-(\Nf+1)/(2\Nf)} \qquad\qquad(\Nf\!\ge\!1)
\quad.
\label{charorig}
\eeq
One now assumes that, in the limit of infinite $V$,
one can define a density of eigenvalues per unit volume, $\rho(\lambda)$.
If so, then one has approximately that
\beq
\rho(\la_c)
\approx {1\over \lambda_c V}
\propto
\la_c^{(\Nf-1)/(\Nf+1)}\qquad(\Nf\!\ge\!1)
\;.
\label{specorig}
\eeq
Since this holds for all large $V$, one can read off the
small $\lambda$ dependence of $\rho(\lambda)$.
One can check this result by noting that it reproduces
the initial result (\ref{sumrorig}):
\beq
\<\sum_n {1\ovr\la_n^2}\> \approx 
\int_{\lambda_c}^{\lambda_{\rm max}} V \rho(\lambda) {1\ovr\la^2} 
\sim V \lambda_c^{-2/(N_f+1)} \propto V^{(N_f+1)/N_f}
\;.
\eeq
Here we approximate finite volume by the lower cut-off
on the $\lambda$ integral.
%Inserting this in the spectral representation of the condensate Smilga finds
%\beq
%\<\psd P_\pm\ps\>\propto
%m^{(\Nf-1)/(\Nf+1)}\,e^{2/(\Nf+1)}\qquad\qquad(m\ll e,\;\Nf\!\ge\!1)
%\;.
%\label{condorig}
%\eeq
%for its mass dependence.

This line of reasoning can be extended to include the
quenched case, $\Nf\!=\!0$.
Smilga notes that if one uses the quenched result for
${\cal G}(x)$, eq.~(\ref{conc6}), one finds that,
asymptotically, the sum rule is
\beq
\<\sum_n {1\ovr\la_n^2}\>\propto
\exp\{V\}\qquad\qquad\qquad(\Nf\!=\!0)
\;,
\label{sumrquen}
\eeq
and so the characteristic size of the lowest eigenvalues is 
exponentially small
\beq
\la_c\propto
\exp\{- V\}\qquad\qquad\qquad(\Nf\!=\!0)
\;.
\label{charquen}
\eeq
We stress that in both these results we control
neither the constant multiplying $V$ in the exponent
(aside from the fact that it is proportional to $e^2$),
nor subleading power dependence multiplying the exponentials.

At this point we extend the results of Smilga by determining
the form of $\rho(\lambda)$ in the quenched theory
\footnote{We acknowledge a useful discussion with A.~Smilga on this point.}.
The same argument used above leads to
\beq
\rho(\la_c)
\approx {1\over \lambda_c V}
\sim {-1\over \lambda_c \log(\lambda_c)}
\;.
\eeq
This cannot be completely correct because it predicts that 
the number of eigenvalues in an interval,
\beq
n(\lambda_2)-n(\lambda_1)=
\int_{\lambda_1}^{\lambda_2} d\lambda\;\rho(\lambda)
= \log[\log\lambda_2/\log\lambda_1] 
\;,
\eeq
diverges as $\lambda_1\to0$.
We propose instead the convergent form
(with missing dimensions supplied by factors of $e$)
\beq
n(\lambda) \propto - {1 \over \log\lambda} 
\qquad \Rightarrow \qquad
\rho(\lambda) \propto {1 \over \lambda (\log\lambda)^2}
\;.
\eeq
This satisfies $n(\lambda_c) \propto 1/V$, which is
an alternative criterion for extracting the infinite volume
eigenvalue density from the finite volume results
\footnote{In the unquenched example above this criterion gives the
same result for $\rho(\lambda)$ as obtained in eq.~(\ref{specorig}).
%%%It is presumably because of the rapid rise in $\rho(\lambda)$ as
%%%$\lambda\to0$ that the extraction of $\rho(\lambda)$ from finite volume
%%%results is more difficult.
}.
In fact, the precise form of $\rho(\lambda)$ is not important:
all that we need is that it falls off as $1/\lambda$, up to
logarithmic corrections, and that these corrections
make it integrable as $\lambda\to0$.

We now insert this result into the expression for the condensate,
approximating the effect of finite volume by a lower cut-off:
\beq
\<\psd P_\pm\ps\> \approx
\int_{\la_c}^\infty d\la\; {2m\ovr\la^2+m^2}\rho(\la)
\;.
\eeq
There are two cases to consider. If $m < \la_c\sim e^{-V}$,
then the IR cut-off is provided by $\la_c$
\beq
\<\psd P_\pm\ps\> \approx
\int_{\la_c}^\infty {2m \over \la^3} \propto {m \over \la_c^2}
\qquad\qquad (m < \la_c)
\;.
\label{condquen1}
\eeq
If $m > \la_c$ then the mass provides the IR cut-off,
and one has
\bea
\<\psd P_\pm\ps\> &\approx&
\int_{\exp(-V)}^m d\la\; {2\ovr m}\rho(\lambda)
+ \int_m^\infty {2m \over \la^3}
\nonumber\\
&=&
{2[n(m)-n(\la_c)]\over m} + {1 \over m}
\nonumber \\
&\sim&
{1 \over m} [1 + O(1/V) ]
\qquad\qquad (m > \la_c)
\;,
\label{condquen2}
\eea
where we have dropped subleading logarithms, which are not controlled.
In other words, the condensate starts out at zero when $m=0$,
as required in finite volume with no chiral symmetry breaking,
rises linearly until $m\simeq\la_c\sim e^{-V}$,
and then falls as $1/m$ up to logarithms.
The point we wish to stress is that eqs.~(\ref{condquen1},~\ref{condquen2}) 
have no IR-divergences.
Sending the volume to infinity at finite $m$, one 
ends up with a finite value. 
If one then sends $m\to0$, however,
the IR-divergence returns as a $1/m$ divergence.

The preceding argument applies for boundary conditions which do not break
chiral symmetry, and thus is not directly applicable to
the massive theory with chirality breaking boundary conditions.
Nevertheless, since a non-zero mass removes the IR divergences with chirally
symmetric BC, it is plausible that it will also do so with chirality
breaking BC. If so, the IR divergences we found
in the mass-derivatives of the condensate must be special to $m=0$.
We conclude this section by presenting a simple model which leads
to this result.

We propose that, when considering the condensate far from the 
chirality breaking boundaries,
their effects can approximately be represented by using
Smilga's analysis but with a small, volume dependent, shift in the mass.
More precisely, we suggest that one can use
eqs.~(\ref{condquen1},~\ref{condquen2}) with
\beq
m \to m_{XB} = m + \la_c \sim m + e^{-V}
\;.
\label{effmass}
\eeq
The shift by $\la_c$ is chosen so as to reproduce the form of
our exact results. In particular, it puts us in the region 
where (\ref{condquen2}) holds, so that
\beq
\<\psd P_\pm \ps\>_{XB}\simeq {1\ovr m_{XB}} = {1\ovr m+\exp(-V)}
\;.
\label{mixedcond}
\eeq
This simple form agrees with all our results and expectations:
when $m=0$ we recover the exponential IR divergence of our exact result;
when $V\to\infty$ at fixed $m$ there is no IR divergence;
and each successive derivative, evaluated at $m=0$, is more IR divergent.
This shows explicitly how the perturbative argument presented above
can be correct but completely misleading.

%%%%%%%%%%%%%%%%%%%%%%%%%%%%%%%%%%%%%%%%%%%%%%%%%%%%%%%%%%%%%%%%%%%%%%%%%%%%%%%

\section{Summary}

%%%%%%%%%%%%%%%%%%%%%%%%%%%%%%%%%%%%%%%%%%%%%%%%%%%%%%%%%%%%%%%%%%%%%%%%%%%%%%%

\noindent
We have presented a study of the chiral condensate in the (massless) quenched
Schwinger model at finite temperature and with bag-inspired spatial boundary
conditions which play the role of a small fermion mass.
Our main analytical result is given in eqs. (\ref{2.3}, \ref{2.4}).

Our first finding is that the (massless) quenched Schwinger model is
ill-defined due to an infrared embarrassment: At finite box-length with the
XB- boundary conditions applied at the two spatial ends, it is a well-defined
theory, but the condensate shows a singular behavior when the boundaries are
sent to infinity.
The precise form of this divergence depends on the details how the volume of
the manifold is sent to infinity, but the generic structure is
``power-law fall-off times exponential divergence''.

Our second observation is that the condensate in the (massless) quenched
Schwinger model does not ``melt'' at high temperatures, 
instead diverging in the limit of infinite temperature.
This has to be contrasted to the case $\Nf\geq1$ where the Schwinger model is
known to show the regular behavior, i.e.\ a condensate which decreases when the
temperature gets large (see e.g. \cite{SaWi,Du} and references therein).

Both predictions can be checked using numerical simulations on a sufficiently
large lattice, and an interesting step in this
direction has recently been taken in Ref.~\cite{KisNar}.

From a conceptual point of view it should be emphasized that our results
for the massless case represent exact analytic findings. They stem from a
textbook-style analytical evaluation of the path integral with fields subject
to the constraints imposed by the boundary conditions. The latter break chiral
symmetry explicitly and prevent exact zero modes of the massless Dirac
operator. This shows that the IR divergence of the condensate in the quenched
massless theory is not tied to a zero in the spectrum of the Dirac operator
on a certain class of gauge field configurations.

On the other hand, our results for the massive theory are more speculative.
By extending the approach of Smilga~\cite{Smilga} we argue that
the quenched condensate is IR finite,
and reconcile this with our finding that the derivatives of the
condensate with respect to the quark mass are IR divergent at $m\!=\!0$.

%%%%%%%%%%%%%%%%%%%%%%%%%%%%%%%%%%%%%%%%%%%%%%%%%%%%%%%%%%%%%%%%%%%%%%%%%%%%%%%

\subsection*{Acknowledgments}
\vspace{-10pt}
\noindent
This work is supported in part by U.S. Department of Energy grant
DE-FG03-96ER40956.
S.D. also acknowledges support from the Swiss National Science Foundation.
Both authors are very grateful to the Center for Computational Physics at the
University of Tsukuba for the hospitality received there.

%%%%%%%%%%%%%%%%%%%%%%%%%%%%%%%%%%%%%%%%%%%%%%%%%%%%%%%%%%%%%%%%%%%%%%%%%%%%%%%

\end{document}